\documentclass[a4paper,11pt]{article}
\usepackage{pos}
\usepackage{graphicx, xcolor, hyperref,soul}
\usepackage[shortlabels]{enumitem}

\newcommand{\cC}{\ensuremath{\mathcal{C}}}
\newcommand{\cH}{\ensuremath{\mathcal{H}}}
\newcommand{\cPT}{\ensuremath{\mathcal{PT}}}
\newcommand{\cCK}{\ensuremath{\mathcal{CK}}}
\newcommand{\cP}{\ensuremath{\mathcal{P}}}
\newcommand{\cT}{\ensuremath{\mathcal{T}}}
\newcommand{\cK}{\ensuremath{\mathcal{K}}}

\newcommand{\Tr}{\text{Tr}}

\DeclareRobustCommand{\eq}[1]{eq.~\eqref{eq:#1}}

\DeclareRobustCommand{\fig}[1]{Fig.~\ref{fig:#1}}

\DeclareRobustCommand{\sec}[1]{Sec.~\ref{sec:#1}}

\DeclareRobustCommand{\refcite}[1]{Ref.~\cite{#1}}

\title{Finite-density QCD, $\mathcal{PT}$ symmetry, and dual algorithms}

\author[a]{Moses A.  Schindler}
\author[b]{Stella T.  Schindler}
\author*[a]{Michael C. Ogilvie}

\affiliation[a]{Department of Physics, Washington University\\
	1 Brookings Drive, St. Louis, MO 63130}

\affiliation[b]{Center for Theoretical Physics, Massachusetts Institute of Technology,\\
	77 Massachusetts Ave., Cambridge, MA 02139}

\emailAdd{mco@wustl.edu}

\abstract{
	Finite-density QCD and many other field theories with sign problems have a $\cPT$-type symmetry.
	After a brief introduction to $\cPT$-symmetric field theories, a real dual representation for $\cPT$-symmetric scalar field theories with complex actions
	is derived. 
	We show that $\cPT$-symmetric field theories  can exhibit exotic behavior, 
	including sinusoidally modulated propagators, disorder lines, and spatially inhomogeneous pattern-forming phases.
	We discuss the interplay of duality, $\cPT$-symmetry and pattern formation using a $\phi^4$ model and $Z(N)$ spin model with sign problems
	as examples. These behaviors may occur in finite-density QCD and related models.
}	
	
\FullConference{%
 The 38th International Symposium on Lattice Field Theory, LATTICE2021
  26th-30th July, 2021
  Zoom/Gather@Massachusetts Institute of Technology
}

\begin{document}
\maketitle

\section{Three key questions}
Determining the phase structure of finite density QCD is an important open problem in lattice gauge theory,
To make progress on this issue, we must address three key, interrelated questions: What is the physical reason that finite-density QCD is a hard problem?
What new phenomena occur at finite density? And how do we know our simulations are getting the physics right?

Physical manifestations of the sign problem are rooted in the non-Hermitian transfer matrix of finite-density QCD.
The transfer matrix is invariant under  $\cCK$ symmetry, where  $\cC$ is charge conjugation and $\cK$ is complex conjugation.
$\cCK$ symmetry is a generalized $\cPT$-type symmetry \cite{Meisinger:2012va}.
$\cPT$-symmetric quantum field theories
can exhibit a wide range of exotic phenomena not seen in models with Hermitian transfer matrices,
such as 
sinusoidally modulated propagators, disorder lines, 
and spatially inhomogeneous ground state field configurations. 
These behaviors may occur in finite-density QCD  \cite{Ogilvie:2018fov, Medina:2019fnx, Schindler:2021otf}.
Related, simpler $\cPT$-symmetric quantum field theories which have these behaviors are natural testing grounds for algorithms such as
the complex Langevin approach \cite{Parisi:1983mgm} and Lefschetz thimbles \cite{Cristoforetti:2012su}.

We begin by introducing basic concepts of $\cPT$ symmetry and their application to finite-density QCD. 
Next, we discuss the variety of different dual forms that a single $\cPT$-QFT may take, and how these forms naturally lead to duality-based algorithms to overcome sign problems. 
Finally, we explain the connection of $\cPT$ symmetry to phenomena such as pattern formation and discuss issues of computational complexity raised by these exotic behaviors. 

\section{Introduction to $\cPT$-symmetric quantum theory}

The development of techniques to understand $\cPT$-symmetric quantum systems was motivated by the Euclidean $i\phi^3$ field theory associated with the Yang-Lee edge singularity \cite{Lee:1952ig}.
The quantum mechanics analogue of $i\phi^3$ theory is the Hamiltonian $\cH= p^2 + ix^3$. 
Although $\cH$ is not Hermitian, it is $\cPT$-symmetric and all of its energy eigenvalues are real \cite{Bender:1998ke}. 
The operator $\cP$ is parity, which acts on $x$ and $p$ as $\cP: x\to -x$ and $\cP:p\to -p$. 
$\cT$ denotes time-reversal, an antilinear operator which acts as $\cT: p \to -p$ and $\cT: i \to -i$, leaving $x$ unchanged.
Let $E$ be an eigenvalue of $H$ with corresponding eigenvector $|E\rangle$.
Noting that $\left[\cPT,H\right]=0$, we see that 
\begin{equation}
H(\cPT)\left|E\right> = (\cPT)H\left|E\right> = (\cPT)E \left|E\right> = E^* (\cPT)\left|E\right>
\end{equation}
so every eigenvalue of $\cH$ must be either real or a complex-conjugate pair. 
This result is true for any system possessing a $\cPT$-type symmetry: we can allow ``$\cP$'' to be any arbitrary linear operator and ``$\cT$'' to be any arbitrary antilinear operator, and the eigenvalues are still ensured to be real or complex conjugates.

If all eigenvalues of a $\cPT$-symmetric system are real, we say that $\cPT$ symmetry is unbroken (a term not directly related to spontaneous symmetry breaking).
Models with an unbroken $\cPT$ symmetry are known to exhibit orthogonal eigenstates, unitary time evolution under a correctly-defined inner product, and other hallmarks of conventional quantum theories \cite{Bender:2007nj, doi:10.1142/q0178}. 
In fact, an entire class of models of the form
\begin{equation}
H=p^2-(ix)^N
\end{equation}
has only real eigenvalues for all $N\geq 2$ \cite{Bender:1998ke}.
Although the $N=3$ model is defined for real $x$, models with higher values of $N$ must be defined via $\cPT$-symmetric
contours in the complex plane. 
These models and their field theory extensions represent good opportunities to cross-check finite-density lattice techniques employing complex contours.

\section{$\cPT$ symmetry in lattice field theory}
The ideas of $\mathcal {PT}$ symmetry in quantum mechanics can be taken over into lattice field theory with a few changes \cite{Meisinger:2012va}.
Rather than work with Hamiltonians, it is conceptually convenient to use the language of the transfer matrix $T=e^{-aH}$,
an operator which implements a discrete time evolution in Euclidean space such that on a lattice of temporal extent $n_t$,
the partition function $Z$ takes the form
\begin{equation}
Z=\text{Tr}(T^{n_t}).
\end{equation}
Schematically, we can obtain the transfer matrix from writing the partition function as a lattice functional integral
\begin{equation}
Z=\int [d\phi]\,e^{-S[\phi]}
\end{equation}
For example, the $i\phi^3$ model with action
\begin{equation}
S=\sum_x \left [{1\over 2}\left(\nabla \phi\right)^2 +ig\phi^3(x)\right]
\end{equation}
is not invariant under the linear transformation $\phi\rightarrow -\phi$ or under complex conjugation, which is antilinear. 
However, it is invariant under the combined operation of both, so we say that the $i\phi^3$
model has a $\mathcal{PT}$-type symmetry.

The transfer matrix $T$ inherits the symmetry properties of $S$, and
many of the results for Hamiltonians in $\cPT$-symmetric quantum mechanics carry over to $\cPT$-symmetric lattice field theories \cite{Meisinger:2012va}. 
In conventional lattice field theories, the transfer matrix is a Hermitian operator, satisfying $T^\dagger=T$, with positive real eigenvalues.
In $\cPT$-symmetric lattice field theories, every eigenvalue of the transfer matrix is either real or part of a complex-conjugate pair.
However, we emphasize that $\cPT$ symmetry does not ensure that all real eigenvalues of $T$ are positive. 
When all eigenvalues of the transfer matrix are real and positive, the system is equivalent to a conventional lattice model under a similarity transformation. 
If some of the eigenvalues of $T$ are negative or complex, a real representation of the transfer matrix still exists, but matrix elements of $T$ in the real representation need \emph{not} be positive.
As we discuss below, this leads to unconventional behaviors like patterned phases and disorder lines.

\section{Finite density QCD has a $\cPT$-type symmetry}

It is well known that QCD with a nonzero chemical potential has a sign problem. 
The sign problem arises from the asymmetric weighting of quark and antiquark worldlines that have nonzero winding around the Euclidean time direction. 
This gives rise to a complex integrand in the functional integral, but leaves behind a $\cPT$-type symmetry, $\cCK$.
Effective models of finite-density QCD, which are commonly studied to gain insight into the full theory, often exhibit both a sign problem and $\cPT$ symmetry. 
Consider a typical Polyakov loop spin model with Euclidean action
\begin{equation}\label{eq:polyakov-spin}
S=-J\sum_{\langle jk\rangle}\left(\Tr\,P_j\,\Tr\,P_k^\dagger +\Tr\,P_j^\dagger\,\Tr\,P_k\right)-H\sum_j\left(e^{\beta\mu}\Tr\,P_j+e^{-\beta\mu}\Tr \,P^\dagger\right).	
\end{equation}
Here, the "spins" are Polyakov loops on spatial sites, taking values in the gauge group $SU(3)$. 
The nearest-neighbor interaction is induced by plaquette terms in the underlying gauge theory action, and the external field interaction is due to heavy quarks at finite density.
It is this second term which becomes complex when $\mu\ne 0$. 
This model constitutes a concrete representation of Svetitsky-Jaffe universality, which maps gauge theories at finite temperature to spin models \cite{Svetitsky:1982gs}.
It is easy to see that \eq{polyakov-spin} has a $\cPT$-type symmetry: 
charge conjugation $\mathcal C$ acts as a linear operator $\mathcal C: P\rightarrow P^*$ on the Polyakov loops, while complex conjugation $\mathcal K$ acts as an antilinear operator $\cK: aP\rightarrow a^*P^*$. 
The action $S$ is invariant under the combined operation $\cCK$.

\section{Simulating $\cPT$-QFTs via a duality transform}\label{sec:simulations}
It is straightforward to transform $\cPT$-symmetric scalar field theories with a complex action into a manifestly real form, using appropriate Fourier transforms of \emph{fields} in the path integral \cite{Ogilvie:2018fov}.
We refer to this transformation as a dual transformation, because it is essentially the first half of a Kramers-Wannier duality transformation. 
In a typical duality transformation, a Fourier transform on an Abelian group $G$ maps lattice variables into a dual group $\tilde G$.
For the group $Z(N)$, the dual group is also $Z(N)$. For the group $U(1)$ the dual group is $\mathbb{Z}$, but in the case of the group $\mathbb{R}$ the dual group is $\mathbb{R}$.

\begin{figure}
\begin{center}
\includegraphics[width=2.5in]{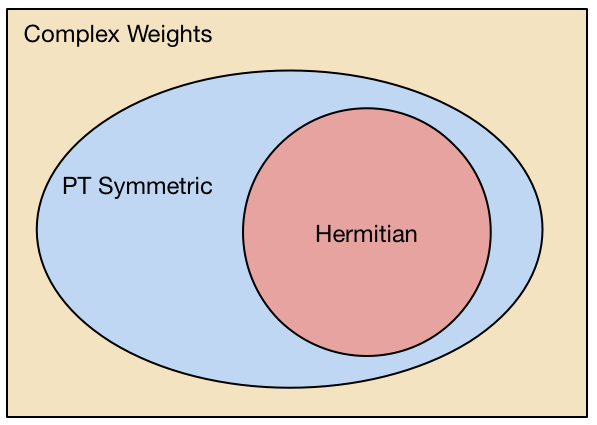}
\end{center}
\caption{Within the space of lattice field theories with complex weights, standard lattice models have weights which are real and positive, while $\cPT$-symmetric models have representations where the weights are real but not necessarily positive.}
\label{fig:ComplexWeights}
\end{figure}

We illustrate the method using a simple model with one field, but it can be easily generalized to models with any number of scalar fields in any dimension. 
Consider the action
\begin{equation}	
	S(\chi)=\sum_{x}\left\{\frac{1}{2}[\partial_{\mu}\chi(x)]^{2}+V[\chi(x)]-ih(x)\chi(x)\right\}
\end{equation}
where $\chi(x)$ is a real field, $h(x)$ is a spacetime-dependent external field and
the potential $V\left(\chi\right)$ obeys the $\cPT$-type symmetry condition
$V(\chi) = V^*(-\chi)$.
The generating function $Z[h]$ is given by the lattice path integral over $\chi(x)$.
Inside that integral, we can transform the kinetic term on each site $x$ using the identity
\begin{equation}
	\exp\left[-\frac{1}{2}\left(\partial_\mu\chi(x)\right)^{2}\right]=\int d\pi_{\mu}(x)\exp\left[\frac{1}{2}\pi_{\mu}(x)^{2}+i\pi_{\mu}(x)\partial_{\mu}\chi(x)\right].
\end{equation}
We can write the potential term on each site as the Fourier transform of a real function $w(\tilde\chi)$:
\begin{equation}
	\exp\left\{-V[\chi(x)]\right\}=\int d\tilde\chi \,\tilde w[\tilde \chi(x)]\, \exp\left[+i\,\tilde\chi(x)\,\chi(x)\right].
\end{equation}
Here, we refer to $\tilde w$ as the \emph{dual weight}, and we say that \emph{dual weight positivity} holds if for all $\tilde\chi$, 
$\tilde w[\tilde\chi(x)] \geq 0$. In that case, we can define a real dual potential $\tilde V\left(\tilde \chi\right)=-\log \tilde w\left(\tilde\chi\right)$.
If we insert the Fourier transform identities for the kinetic and potential terms into the path integral, the integration over $\chi(x)$ can be performed, resulting in a \emph{dual action}
\begin{equation}\label{eq:dual-action}
	\tilde{S}= \sum_{x}\left\{\frac{1}{2}\pi_{\mu}^{2}(x)+\tilde{V}[\partial\cdot\pi(x)-h(x)]\right\}.
\end{equation}
When dual weight positivity holds, the dual path integral has a manifestly positive integrand and it may be simulated by standard lattice methods.
Mean field theory and other conventional analytical methods can also be applied to the dual form.
Expressions for correlation functions of the field $\chi(x)$ are obtained by functional differentiation with respect to $h(x)$.
Dual weight positivity is an important condition singling out a special class of simulatable models, as shown in \fig{ComplexWeights}.

When dual weight positivity is not satisfied, the integrand of the path integral is real but not positive everywhere, so the sign problem is transformed but not eliminated.
However, there may be models in  the same universality class which do not have a sign problem.
One approach is to start from a real dual potential $\tilde{V}$ parametrized to have the desired behavior. 
A second approach, inspired by the Migdal-Kadanoff real-space renormalization group transformation  \cite{Migdal:1975zg, Migdal:1975zf,Kadanoff:1976jb}, is to simply square dual weight $\tilde w$ of the original model.
Because $\tilde w$ is real, $\tilde w^2$ is positive, and the resulting model should be in the same universality class as $\tilde w$.
We conjecture that each universality class contains at least one model with a manifestly positive representation which can be simulated by standard methods.

\section{Dual forms of $\cPT$-symmetric QFTs}

$\cPT$-symmetric quantum field theories can have multiple equivalent forms.
We can explore some of these forms for the case of a Euclidean action with two scalar fields:
\begin{equation}\label{eq:two-fields}
	\mathcal{L}(\phi,\chi) = \frac{1}{2}(\nabla_\mu \phi)^2 + \frac{1}{2}(\nabla_\mu \chi)^2+  \frac{1}{2}m_\chi^2 \chi^2   -ig\phi\chi+\lambda(\phi^{2}-v^{2})^{2}+h\phi.
\end{equation}
This $\cPT$-symmetric extension of Hermitian $\phi^4$ theory was first studied in \refcite{Ogilvie:2018fov} and its phase diagram was extensively analyzed in \refcite{Medina:2019fnx}. 
It is possible to cast \eq{two-fields} into a real local form using the method in \sec{simulations} by the dual transform method discussed above:
\begin{equation}
	\tilde{S} = \sum_{x} \left[\frac{1}{2}(\nabla_\mu\phi)^2 + \frac{1}{2}\pi_\mu^2  + \frac{1}{2m_\chi^2}(\nabla\cdot \pi - g\phi)^2 + \lambda(\phi^2-v^2)^2 + h\phi \right].
\end{equation}
Simulations of this model display exotic behaviors characteristic of $\cPT$-symmetric systems: inhomogeneous phases, disorder lines, and sinusoidally modulated propagators \cite{Medina:2019fnx, Schindler:2021otf}. 
We can also recast \eq{two-fields} into a real \emph{nonlocal} form by integrating out the $\chi$ field from \eq{two-fields}
\begin{equation}
	S_{\text{eff}}=\sum_{x} \left[\frac{1}{2}(\partial_{\mu}\phi(x))^{2}+\lambda(\phi^{2}-v^{2})^{2}+h\phi\right]
	+\frac{g^{2}}{2}\sum_{x,y}\phi(x)\Delta(x-y)\phi(y),
\end{equation}
where $\Delta(x)$ is the propagator for $\chi$. 
The model studied in \cite{Medina:2019fnx, Schindler:2021otf} could be called a Yukawa-frustrated $\phi^4$ model.
In the limit $m_\chi\rightarrow 0$, this form of the action is known as the Coulomb-frustrated $\phi^4$ model, and has been studied as a pattern-forming model for some time in condensed matter physics.
In this form, it is clear that the $\chi$ field acts to restore the $Z(2)$ symmetry, consistent with a common picture of pattern formation: resulting as a consequence of opposing attractive and repulsive forces.
For a more detailed analytic and numeric description of pattern formation, see \refcite{Schindler:2021cke}. 

Finally, it is known that $\cPT$-symmetric systems have deep connections to higher-derivative theories \cite{Bender:2007wu}. 
We can see this connection in \eq{two-fields} by simply expanding the nonlocal interactions induced by $\chi$ into higher-derivative terms, to arrive at an approximate, effective real form: 
\begin{equation}
	S_{\text{eff}}\approx\sum_{x} \left[\frac{1}{2}(\partial_{\mu}\phi(x))^{2}+\lambda(\phi^{2}-v^{2})^{2}+h\phi\right]
	+\frac{g^{2}}{2m_\chi^2}\sum_{x}\left[\phi(x)^2-{1 \over m_\chi^2}(\partial_{\mu}\phi(x))^{2}+\dots\right].
\end{equation}
As $g$ increases from zero, the coefficient of the quadratic kinetic term decreases, leading to a
Lifshitz instability where the homogeneous phase becomes unstable with respect to sinusoidal modulations
at a particular wavenumber, resulting in a stable inhomogeneous ground state. This is precisely what is seen in lattice simulations of the model.

\section{$Z(N)$ spin models and pattern formation}

$Z(N)$ spin models with a non-zero chemical potential also have a  $\mathcal{PT}$-type symmetry and 
 are closely related to finite-density QCD.
Consider a simple $Z(N)$ spin model in $d$ dimensions. The standard nearest-neighbor
action for such a model is
\begin{equation}
	{\mathcal H}=-{J \over 2}\sum_{\left< j\nu \right>}\left(z_j z_{j+\hat\nu}^*+z_j^* z_{j+\hat\nu}\right)
\end{equation}
where $z_j$ is a site-based variable in $Z(N)$. This model has a Hermitian transfer matrix,
but we can make it complex by introducing a chemical potential into the nearest-neighbor interaction
in the $d$ direction:
\begin{equation}
	z_j z_{j+\hat d}^*+  z_j^* z_{j+\hat d} \rightarrow
e^\mu z_j z_{j+\hat d}^*+ e^{-\mu}z_j^* z_{j+\hat d}.
\end{equation}
This model can be viewed as a simplified model of quarks in a chemical potential.
It is $\mathcal{PT}$-symmetric and the dual weights are real.
Pattern formation occurs in the dual representation of this model
where the nearest-neighbor interaction
in the $d$ direction has the form
\begin{equation}
	z_j z_{j+\hat d}^*+  z_j^* z_{j+\hat d} \rightarrow
	e^{i\mu} z_j z_{j+\hat d}^*+ e^{-i\mu}z_j^* z_{j+\hat d}.
\end{equation}
This model has been called the $Z(N)$ chiral model because a nonzero $\mu$ favors
the angle between adjacent spins in the $\hat d$ direction differing by $\mu$.
Thus the state of lowest energy may be one where the spins twist to the left or right along the
$d$ direction.
The phase diagram of this model in the $J$-$\mu$ plane reveals a so-called "Devil's Flower"
with a rich set of patterned behaviors \cite{YEOMANS19841}. 

These two models are not equivalent under duality, but they belong to classes of models which
map into each other under duality: nearest-neighbor $\mathcal{PT}$ $Z(N)$ models with real and
complex actions. 
Only in the case of the $Z(N)$ model with a Villain action is there a simple duality \cite{Meisinger:2013zfa, Meisinger:2013zba}:
\begin{align}
J\rightarrow \tilde J= {N^2 \over 4\pi^2 J}
\mu\rightarrow \tilde \mu= -{2\pi i J\mu \over N}
\end{align} 
In two dimensions, duality maps spin systems to spin systems, but the map
is between spin and gauge systems in three dimensions, and gauge to gauge
in four dimensions.
The two-dimensional case has a nice field-theoretic interpretation: it is
the universality class of two-dimensional  $Z(N)$ parafermions 
and the patterned behavior in the chiral model corresponds to states with nonzero $N$-ality realized as kinks.

\section{Implications for computational complexity}
One might be tempted to assume that by recasting a model from a form with an ostensible sign problem into a manifestly real positive form, we have fully circumvented
all issues of computational complexity.
It is far from clear that this is the case.
Simulation of a system in an inhomogeneous equilibrium state presents additional issues.
Patterned phases may have a large number of near-zero modes
associated with moving the boundaries of domains, leading to long equilibration times, both in physical systems and in simulations.
It has been proposed that scalar field theory models with long-range interactions \cite{PhysRevLett.85.836} or higher-derivative interactions \cite{WU20021} can model glassy behavior, a prototypical non-deterministic non-polynomial hard (NP-hard) problem.
This is notable, because these models do not have frustration introduced by quenched bond or site disorder, unlike
a spin glass.
On the other hand, the well-known work of Troyer and Wiese \cite{Troyer:2004ge} shows that the sign problem of fermionic many-body systems is NP-hard by demonstrating its equivalence to finding the ground state of a random-bond Ising model.
In either case, computational complexity has its origins in the inhomogeneity of ground states and equilibrium states,
for which we lack a complete understanding.

\section{Conclusions}

Many lattice models with sign problems have an underlying $\cPT$-type symmetry, with finite-density QCD and related models an important class.
$\mathcal{PT}$-symmetric lattice models have a richer structure than their conventional counterparts.
Spectral positivity is lost, and exotic phases may emerge. 
Our recent work on the Yukawa-frustrated $\phi^4$ model \cite{Medina:2019fnx}
as well as a heavy-fermion QCD-like effective model \cite{Schindler:2021otf}
indicate real progress can be made.
We believe that a number of interesting and important models
can now be treated, including
models in the $i\phi^3$ universality class,
non-unitary minimal conformal models \cite{Dotsenko:1984nm},
affine Toda models \cite{Hollowood:1991vfe},
and $\mathcal{PT}$-symmetric $SU(N)$ models \cite{Fring:2020bvr,Fring:2020xpi, Fring:2020wrj, Fring:2021zci}.

\section*{Acknowledgments}
STS was supported by a Graduate Research Fellowship from the U.S. National Science
Foundation under Grant No. 1745302; the U.S. Department of Energy, Office of Science, Office
of Nuclear Physics, from DE-SC0011090; and fellowship funding from the MIT Department of
Physics.

\bibliographystyle{JHEP}
\bibliography{ogilvie-edits}

\end{document}